# Focus Plus: Detect Learner's Distraction by Web Camera in Distance Teaching

Eason Chen[1 *], Yuen Hsien Tseng[2], Kuo-Ping Lo[3]

National Taiwan Normal University

eason.tw.chen@gmail.com

## Abstract

Distance teaching has become popular these years because of the COVID-19 epidemic. However, both students and teachers face several challenges in distance teaching, like being easy to distract. We proposed Focus+, a system designed to detect learners' status with the latest AI technology from their web camera to solve such challenges. By doing so, teachers can know students' status, and students can regulate their learning experience. In this research, we will discuss the expected model's design for training and evaluating the AI detection model of Focus+.

## 1. Research Purpose

Distance teaching has become popular these years because of the COVID-19 epidemic. As a result, teachers nowadays teach their students synchronously via meeting software like Google Meet, Zoom, WebEx, etc. However, students face several challenges, primarily the high chances of distraction by external sources, like noise, or multitasking, like refreshing social media app (Xiao & Wang, 2017). Moreover, lecturers also feel stressed because they can't see their students.

In this research, we proposed an AI model and system, that can detect learner's emotion and anomaly level on their localhost. Our ultimate goal is to solve the problems mention above, by developing a platform called 'Focus+ (Spell: Focus Plus)'. Focus+ aims to improve students' concentration by logging their facial emotion and concentration status for self-regulated learning. Moreover, Focus+ allows lecturers to remotely check students' learning status without violating their privacy by receiving processed data. By doing so, we hope Focus+ can improve the distance teaching experience for both students and teachers, and create a new source of education big data.

## 2. Research background

**Distraction**

Medina (2011) indicated that people's attention is best during the first 10 minutes. After that, they will start to distract. By definition, there are two types of distraction, divided attention and mind wandering. Divided attention (DA) occurs when simultaneous stimulation divides attention resources, then influence people's performance(Kahneman, 1973). DA often happened when intended multi-tasking (internal distraction) or external distraction. On the other hand, mind wandering (MW) is independent of stimulation. MW happens by changing attention from task-related thoughts to unrelated thoughts(Mason et al., 2007). In this research, we will try to cover both types of distraction.

**The measure of distraction**

Researchers have tried to use different physiological signals to infer subjects' attention status, such as Electroencephalography (EEG)(Rodrigue, Son, Giesbrecht, Turk, & Höllerer, 2015), eye tracking(Rodrigue et al., 2015), and heart rate(Xiao & Wang, 2017). Nevertheless, students need to have special equipment to detect these signals in practice. In this research, we will use a device that every laptop has: web camera.

**Measure Facial Expression by AI model**

Kartynnik, Ablavatski, Grishchenko, and Grundmann (2019) designed a model: MediaPipe Facemesh, which can detect 3D facial landmarks via lightweight devices such as mobile-GPU or CPU from web camera. This model can detect humans' faces in an image and transform them into 486 mesh key points in a 3D coordinate space (x, y, z). These feature vector can be as input of other model. For example, Ayache and Alti (2020) indicated that with the input of the facial landmarks' vector, the performance of the facial emotion classification model is very accurate.

Based on the 3D coordinate of a human face, Kummen et al. (2021) proposed an open-source library that includes eye tracking module that can calculate the eye gaze vector.

**Detect Distraction**

To detect distraction, we assume people show somehow different when they focus and when they distract. Therefore, we only need people's focus data as training data. Then we can use the anomaly detection method to detect distraction. To run a model in the front-end, we will use One Class SVM as an anomaly detector (Schölkopf, Williamson, Smola, Shawe-Taylor, & Platt, 1999) via LIBSVM library (Chang, 2011).

## 3. Methodology

We will use the AI model (see Figure 1) to reach the following two goals:
1. Detect learners' emotion.
2. Detect learners' anomaly level.

Focus+ will use the image captured by the web camera, MediaPipe Facemesh (Kartynnik et al., 2019) will use them to detect learners' face landmarks.

After that, Focus+ will send these points into a Neuro Network Classification Model to predict the emotion. This classification model will train by Affectnet dataset (Mollahosseini, Hasani, & Mahoor, 2017), which contain over 42000 labeled human face images and its' emotion. The emotion includes Neutral, Happiness, Sadness, Surprise, Fear, Disgust, Anger, Contempt, None, Uncertain, and No-Face.

Next, Focus+ will use a One Class SVM model to infer learners' anomaly status. The model will use the following metrics as the input:
1. Learner's emotions by emotion classification model.
2. Eye gaze vector predict by module by Kummen et al. (2021).
3. The angle of the face facing based on the MediaPipe Facemesh by Kartynnik et al. (2019).
4. The face landmarks' vector after reducing dimensionality.
5. The other metadata, such as current course type.

Finally, the One Class SVM will output the anomaly level in the range of 0 to 1.

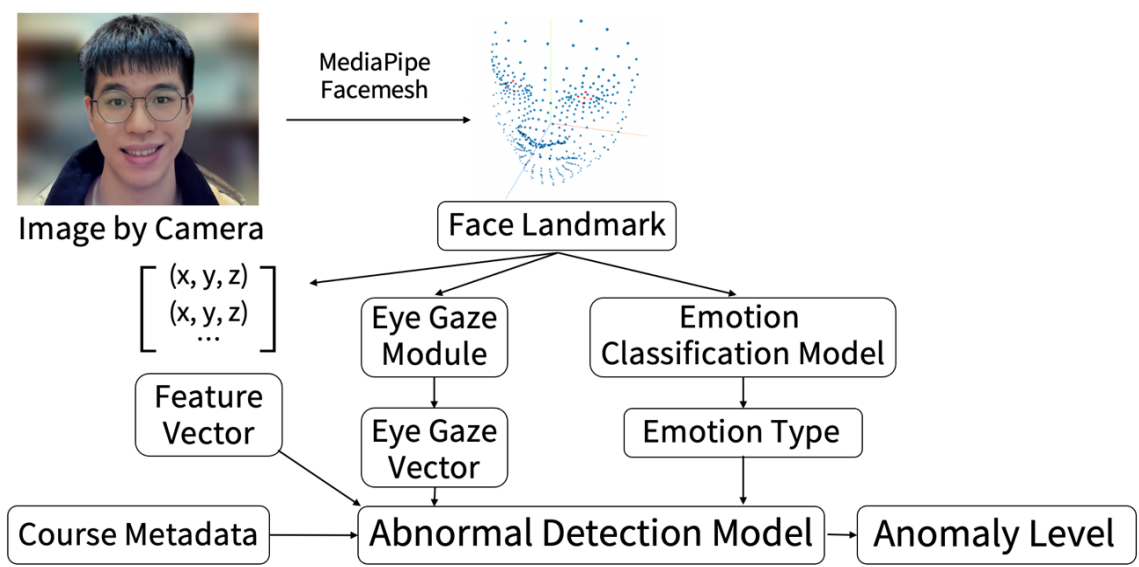

Figure 1. The data flow of the AI model

**How to train the model**

About the training metrics and label, because each user has a different facial expression and learning environment, Focus+ will use the Active Learning strategy. Active Learning is that the pre-train model will retrain and improve its accuracy based on the feedback of the deploy environment(Settles, 2009). To be specific, when setting up the Focus+, each user will need to follow a series of instructions, such as looking at a specific point on the screen, to fine-tune the eye gaze vector on the screen.

To create training data for anomaly level. Since attention is best during the first 10 minutes (Medina, 2011), the system will use the first 10 minutes as training data. Then, we will use LIBSVM (Chang, 2011) to train a One Class SVM as an anomaly detector. If the input is somehow different from most training data, the model will predict it as anomaly.

**Evaluation Method**

We will invite 15 testers to evaluate Focus+. The test consisted of the following three parts:

1. Introduction: we will introduce the purpose of this test. And we will let the participants fine-tune their model by web camera.

2. Then, the participants will work on test sessions A, B, C: The system will randomly assign both these three sessions' orders and video content. After each test session, participants will do a ten-question quiz (Multiple choice) about the content of the video and report their perceived distraction level status on a 7-point Likert scale.

A. Focus Session (FS): Participants watch ten minutes of video without any distraction.

B. Divided Attention Session (DAS): Participants watch ten minutes of video but are required to count how many notifications sounds pop in the background.

C. Mind Wandering Session (MWS): Participants watch ten minutes of video but must think unrelated stuff, like the top ten restaurants he likes.

1. Evaluate: we will then present their Focus log on each session. Participants will rate the AI model's accuracy in each session on a 7-point Likert scale. Then they will fill the SUS (Brooke, 1996) and Qualitative feedback about their experience and expectation of Focus+.

**Current Progress**

Currently, we have built a Proof of Concept (PoC) version that can detect the anomaly level by the learner's face direction (Fig. 2, 3). The next version of Focus+ which uses the model design above is underbuilding. We will demo Focus+ at the presentation day to get feedbacks if allowed.

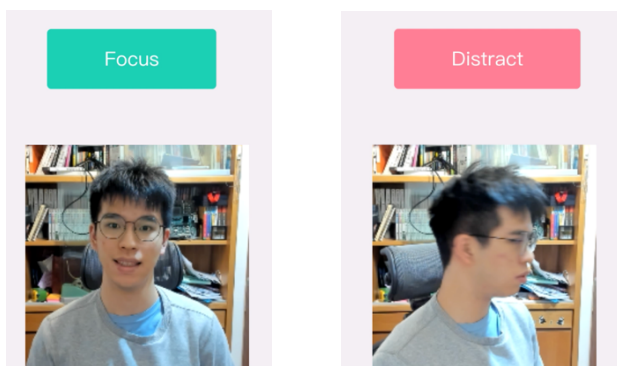
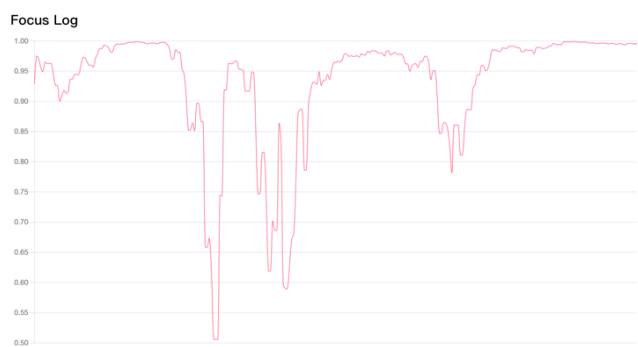

Fig. 2. PoC Model: Focus: face center, Distract: face aside

Fig. 3. Students' Focus Log based on the PoC model

## 4. Expected Results

**Expected result from the evaluation**

We expect the following result:
1. Participants' emotion frequency and mean of anomaly level is different between FS, DAS, MWS.
2. FS have lower anomaly level mean than DAS and MWS.
3. At DAS, participants' anomaly will rise when the notification sound appear.
4. Participants' self report on each session related to their anomaly level.
5. User like Focus+ and expect to use them in their daily life.

**Expect Data Analyze Method**

We will use SPSS 23.0 to analyze. We will use One-Way ANOVA to find difference anomaly level in each session within each user. Then use Scheffe's Post Hoc to check the different between FS, DAS and MWS. And we will use chi-squared test to identify the different frequency of the emotion type in each session.

**Expected User Flow for Focus+ (Fig. 4.)**

To use Focus+, students need to login to the platform and open their camera. After that, Focus+ will compute their emotion and anomaly level at their local CPU and send the metrics back to the server. Then, the teacher can see students' metrics on the real-time dashboard. After the class, students will receive the visualized log for their distraction level and emotion type to do self-regulated learning. Focus+ also provide instructions for self-regulated learning for reference.

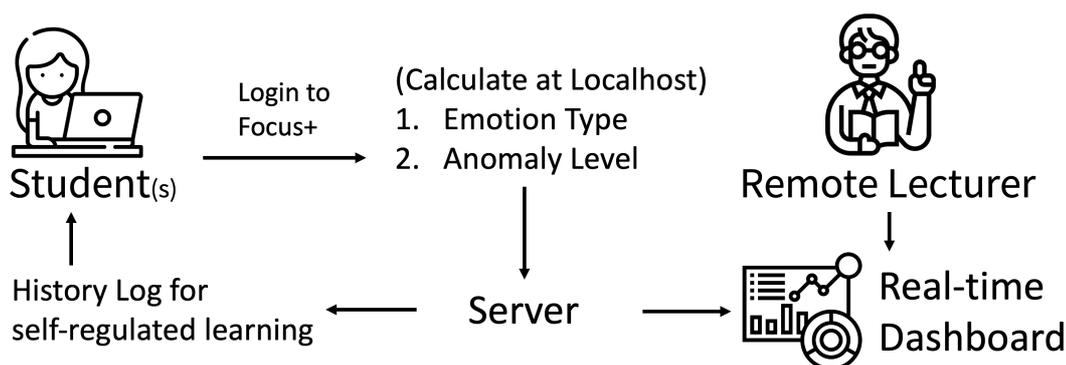

Fig. 4. The user flow of the Focus+

## 5. References

Ayache, F., & Alti, A. (2020). Performance Evaluation of Machine Learning for Recognizing Human Facial Emotions. *Rev. d'Intelligence Artif., 34*(3), 267-275.

Brooke, J. (1996). SUS-A quick and dirty usability scale. *Usability evaluation in industry, 189*(194), 4-7.

Chang, C.-C. (2011). " LIBSVM: a library for support vector machines," ACM Transactions on Intelligent Systems and Technology, 2: 27: 1--27: 27, 2011. *http://www. csie. ntu. edu. tw/~ cjlin/libsvm, 2*.

Kahneman, D. (1973). *Attention and effort* (Vol. 1063): Citeseer.

Kartynnik, Y., Ablavatski, A., Grishchenko, I., & Grundmann, M. (2019). Real-time facial surface geometry from monocular video on mobile GPUs. *arXiv preprint arXiv:1907.06724*.

Kummen, A., Li, G., Hassan, A., Ganeva, T., Lu, Q., Shaw, R., . . . Almazov, E. (2021). MotionInput v2. 0 supporting DirectX: A modular library of open-source gesture-based machine learning and computer vision methods for interacting and controlling existing software with a webcam. *arXiv preprint arXiv:2108.04357*.

Mason, M. F., Norton, M. I., Van Horn, J. D., Wegner, D. M., Grafton, S. T., & Macrae, C. N. (2007). Wandering minds: the default network and stimulus-independent thought. *Science, 315*(5810), 393-395.

Medina, J. (2011). *Brain rules: 12 principles for surviving and thriving at work, home, and school*: ReadHowYouWant. com.

Mollahosseini, A., Hasani, B., & Mahoor, M. H. (2017). Affectnet: A database for facial expression, valence, and arousal computing in the wild. *IEEE Transactions on Affective Computing, 10*(1), 18-31.

Rodrigue, M., Son, J., Giesbrecht, B., Turk, M., & Höllerer, T. (2015). *Spatio-temporal detection of divided attention in reading applications using EEG and eye tracking.* Paper presented at the Proceedings of the 20th international conference on intelligent user interfaces.

Schölkopf, B., Williamson, R. C., Smola, A. J., Shawe-Taylor, J., & Platt, J. C. (1999). *Support vector method for novelty detection.* Paper presented at the NIPS.

Settles, B. (2009). Active learning literature survey.

Xiao, X., & Wang, J. (2017). *Understanding and detecting divided attention in mobile mooc learning.* Paper presented at the Proceedings of the 2017 CHI conference on human factors in computing systems.